\documentclass[journal]{IEEEtran}

\ifCLASSINFOpdf
\else
   \usepackage[dvips]{graphicx}
\fi
\usepackage{url}

\usepackage{graphicx}
\usepackage{amsmath}
\usepackage{enumerate}
\usepackage[table,xcdraw]{xcolor}
\usepackage{flushend}
\usepackage{multirow}
\usepackage{multicol}

\begin{document}

\title{Significance of Chirp MFCC as a Feature in Speech and Audio Applications}

\author{S. Johanan Joysingh, P. Vijayalakshmi, and T. Nagarajan
\thanks{S. Johanan Joysingh is with Sri Sivasubramaniya Nadar College of Engineering, Kalavakkam, Chennai, India. (e-mail: johananjoysinghs@ssn.edu.in).}
\thanks{P. Vijayalakshmi is with Sri Sivasubramaniya Nadar College of Engineering, Kalavakkam, Chennai, India. (e-mail: vijayalakshmip@ssn.edu.in).}
\thanks{T. Nagarajan is with Shiv Nadar University Chennai, Kalavakkam, Chennai, India (e-mail: nagarajant@snuchennai.edu.in).}}

\maketitle

\begin{abstract}
A novel feature, based on the chirp z-transform, that offers an improved representation of the underlying true spectrum is proposed.
This feature, the chirp MFCC, is derived by computing the Mel frequency cepstral coefficients from the chirp magnitude spectrum, instead of the Fourier transform magnitude spectrum.
The theoretical foundations for the proposal, and the experimental validation using product of likelihood Gaussians, to show the improved class separation offered by the proposed chirp MFCC, when compared with vanilla MFCC are discussed.
Further, real world evaluation of the feature is performed using three diverse tasks, namely, speech-music classification, speaker identification, and speech commands recognition. 
It is shown in all three tasks that the proposed chirp MFCC offers considerable improvements.
\end{abstract}

\begin{IEEEkeywords}
z-transform, Fourier transform, chirp spectrum, MFCC
\end{IEEEkeywords}

\IEEEpeerreviewmaketitle

\section{Introduction}

\IEEEPARstart{F}{eatures} must capture essential and distinguishing characteristics of a class of signals accurately. 
In essence they are a conversion of these characteristics into values that can be mathematically manipulated.
This work focuses on a feature for audio signals.
There are two broad approaches by which features can be derived.
The first method is to hand-craft features, whereby, the values corresponding to diverse and prominent characteristics of the signal are derived individually, and combined together to form a feature vector. 
In the second method, we aim to provide an accurate and succinct representation of the signal, from which the machine learning algorithm finds underlying patterns.
These include various spectral features such as the magnitude spectrum, power spectral density, phase spectrum, etc., 
and the filtered spectral features such as Mel frequency cepstral coefficients (MFCC), linear predictive cepstral coefficients (LPCC), etc.

Many task-specific and generic features have been proposed in the literature for speech and audio applications. 
The examples include features based on, 
multi-rate signal processing and the Teager energy operator \cite{jabloun1999teager}, 
Mel-frequency discrete wavelet coefficients \cite{tufekci2000feature}, 
multi-resolution cochleagram \cite{chen2014feature}, 
minimum variance distortionless response spectrum \cite{yapanel2003new}, 
power normalized cepstral coefficients \cite{kim2009feature}, and
modified group delay \cite{hegde2007significance}.
MFCC \cite{davis1980comparison} is ubiquitous in speech and audio technology, and is used in applications such as: automatic speech recognition, language identification
\cite{muthusamy1994automatic} \cite{zissman1996comparison},
music genre classification
\cite{lee2007automatic} \cite{lee2009automatic}, 
speaker identification \cite{nakagawa2011speaker} \cite{zhao2013analyzing}, etc.
In \cite{zheng2001comparison}, different implementations of the MFCC, based on modification of the number, shape and spacing of the filters, and the warping method, are explored.
The current work presents a feature based on the MFCC and the chirp spectrum, that can be used for a variety of audio-centric applications.

The discrete Fourier transform (DFT), the first step in computing the MFCC, can be understood as a special case of the Z-transform. 
The Z-transform is defined as,
\begin{equation}
    X(z) = \sum_{n=0}^{N-1} x[n] z^{-n},
\label{eq_ztransform_1}
\end{equation}
where $x[n]$ is the input signal of size $N$. 
The discrete-time Fourier transform computed at a particular radius can be given as, 
\begin{equation}
    X(z)|_{z=re^{j\omega}} = \sum_{n=0}^{N-1} x[n] (re^{j \omega})^{-n},
\label{eq_ztransform_2}
\end{equation}
where $r$ is the radius, and $\omega$ the angular frequency.
DTFT is used here, for convenience

Generally, when $r\neq1$, it is called the chirp spectrum \cite{rabiner1969chirp} \cite{rabiner2004chirp}.
From another perspective,
eq. \ref{eq_ztransform_2} offers basis functions, or sinusoids, 
that are exponentially growing, exponentially decaying or sustained in nature, depending on the value of the radius $r$. 
A special case here is the z-transform measured at the unit circle ($r=1$, sustained sinusoid), which is equivalent to the DTFT.
Hence, the set of basis functions to compute the chirp spectrum comprises exponentially growing and decaying sinusoids only. 

In this work, we propose a spectral feature --- the chirp MFCC, which is the MFCC computed from the chirp magnitude spectrum.
The proposed feature provides an improved representation of the signal.
To this end, the manuscript is organised as follows. 
In Section \ref{sec:analysis}, the theoretical basis and experimental validation of the proposal are provided in detail.
In Section \ref{sec:evaluation}, three classification tasks, namely speech-music classification, speaker identification, and speech commands recognition are discussed. 
Comparison of the proposed chirp MFCC and vanilla MFCC, as a feature for these tasks, is discussed in this section. 

\section{Analysis}
\label{sec:analysis}

In this section, the motivation for estimating the chirp spectrum, instead of the discrete-time Fourier transform, is first demonstrated through two analyses:
\begin{enumerate}
    \item the theoretical analysis of the estimation of phase using DTFT versus the chirp spectrum, for a single-pole system.
    \item the empirical analysis of the estimation of phase using DTFT versus the chirp spectrum, for a multi-pole system.
\end{enumerate}
Further in Section.\ref{sec:analysis_real_speech}, the best radius for chirp spectrum estimation for real speech is estimated through linear predictive analysis.
Following which, in Section. \ref{sec:experimental_analysis}, an experimental proof of the improved class separation offered by chirp spectrum is demonstrated using product of likelihood Gaussians.

\subsection{Analysis of Single-Pole System}
\label{sec:one_pole_system}
Consider a causal sequence $x(n)$ generated by a single pole system. 
Let the discrete-time Fourier transform of the causal sequence $x(n)$ be,
\begin{equation}
\label{eq:ft_long}
    X(\omega)=\sum_{n=0}^{N-1}x(n)cos(\omega n)
    - j\sum_{n=0}^{N-1}x(n)sin(\omega n).
\end{equation}

\noindent With reference to eq.\ref{eq_ztransform_2}, the basis sequence used for the transform in eq.\ref{eq:ft_long} is $e^{-j\omega n}$, where $r = 1$. 
The corresponding phase spectrum $\theta(\omega)$ can be given as, 

\begin{equation}
    \theta(\omega) = tan^{-1} 
    \bigg(\frac{\sum_{n=0}^{N-1} x(n) sin(\omega n)}
    {\sum_{n=0}^{N-1} x(n) cos(\omega n)} \bigg)
\end{equation}

The following three cases illustrate the computation of 
the value of the phase at $\omega = \omega_0$ 
for different input and basis sequences:

Case 1: For an input sequence $x(n)$, in the form of
a \textit{sustained} oscillation, with frequency $\omega_0$
and phase $\phi$, given by, 
\begin{equation}
\label{eq:case1_input}
    x(n) = cos(\omega_0 n + \phi)
\end{equation}

the phase at $\omega = \omega_0$ can be given as,

\begin{equation}
\label{eq:case1_1}
\begin{split}
    \theta(\omega_0) 
    & = tan^{-1} \bigg( \frac
    {\sum_{n=0}^{N-1} cos(\omega_0 n + \phi) sin(\omega_0 n)}
    {\sum_{n=0}^{N-1} cos(\omega_0 n + \phi) cos(\omega_0 n)}
    \bigg) \\
    & = tan^{-1} \bigg(\frac
    {\sum_{n=0}^{N-1} cos(\omega_0 n + \phi) cos(\omega_0 n - 90)}
    {\sum_{n=0}^{N-1} cos(\omega_0 n + \phi) cos(\omega_0 n)}
    \bigg) 
\end{split}
\end{equation}

Considering the two terms in the numerator (and the denominator) as the dot product of two vectors of unit magnitude, 
eq.\ref{eq:case1_1} can be \textit{simplified} to, 
\begin{equation}
    \label{eq_simplified_phase}
    \theta(\omega_0) = tan^{-1}(tan\ \phi) = \phi,
\end{equation}
which is the initial phase of the input sequence in eq.\ref{eq:case1_input}.

Case 2: For an input sequence in the form of
a \textit{decaying} oscillation $x_d(n)$,
with frequency $\omega_0$, phase $\phi$, and
decaying at the rate of $a^n$, for $0 < a < 1$,
given by,

\begin{equation}
    \label{eq_input_seq2}
    x_d(n) = a^n cos(\omega_0 n + \phi)
\end{equation}

the phase at $\omega = \omega_0$ is given as,

\begin{equation}
\label{eq_phase_ip2}
    \theta(\omega_0) = tan^{-1} 
    \bigg(\frac
    {\sum_{n=0}^{N-1} a^n cos(\omega_0 n + \phi) cos(\omega_0 n - 90)}
    {\sum_{n=0}^{N-1} a^n cos(\omega_0 n + \phi) cos(\omega_0 n)}
    \bigg)
\end{equation}

It can be seen that,
eq. \ref{eq_phase_ip2} \textit{cannot be simplified} to
eq. \ref{eq_simplified_phase}, and hence
the estimated phase at $\omega_0$ can be given as,
\begin{equation}
\label{eq_phase_error}
    \theta_e(\omega_0) = \phi + \epsilon,
\end{equation}
where $\epsilon$ is the error in phase estimation. 
It can also be inferred from eq. \ref{eq_phase_ip2} that this is because of the decaying component $a^n$ in the input sequence.

Since both the magnitude and the phase are computed from the real and imaginary coefficients of the discrete-time Fourier transform, these errors are propagated to both the magnitude and the phase spectra.


Case 3: For an input sequence in the form of a
\textit{decaying} oscillation $x_d(n)$ as given by eq. \ref{eq_input_seq2}, the error in phase estimation, at a particular frequency $\omega = \omega_0$, can be nullified under the following two conditions. 
\begin{enumerate}
    \item It can be readily seen from eq. \ref{eq_phase_ip2} that if the input sequence is explicitly weighted by $a^{-n}$, the phase at $\omega_0$ can be given in the \textit{simplified} form as in eq. \ref{eq_simplified_phase}. 
    \item The same can also be achieved if the value of the radius $r$, that is used to compute the chirp spectrum, as in eq.\ref{eq_ztransform_2}, is set to $r = a$.
    Here, weighting the basis sequence implicitly nullifies the decaying component of the original signal.
\end{enumerate}
From this analysis we can conclude that the phase error for a system with one decaying sinusoidal component is zero when the chirp spectrum is computed at a radius at which the sinusoidal component was generated. 
Computing the precise value of the radius $r$ is not straight-forward when there are multiple sinusoidal components in a synthetic signal, or in the case of real signals.
The following two sections discuss the ideal value of radius $r$ in these two cases.

\subsection{Analysis of Multi-Pole Systems}
\label{sec:multiple_pole_systems}
To estimate the error in phase estimation when multiple frequency components are present in the input sequence, experiments are carried out with systems that have two, three and four poles, and empirical observations are made.
The observations made for four pole systems were reflected in the systems with two and three poles as well, hence only the former is described below.

\subsubsection{Experimental Setup}
\begin{figure}[ht]
    \centering
    \includegraphics[width=0.45\textwidth]{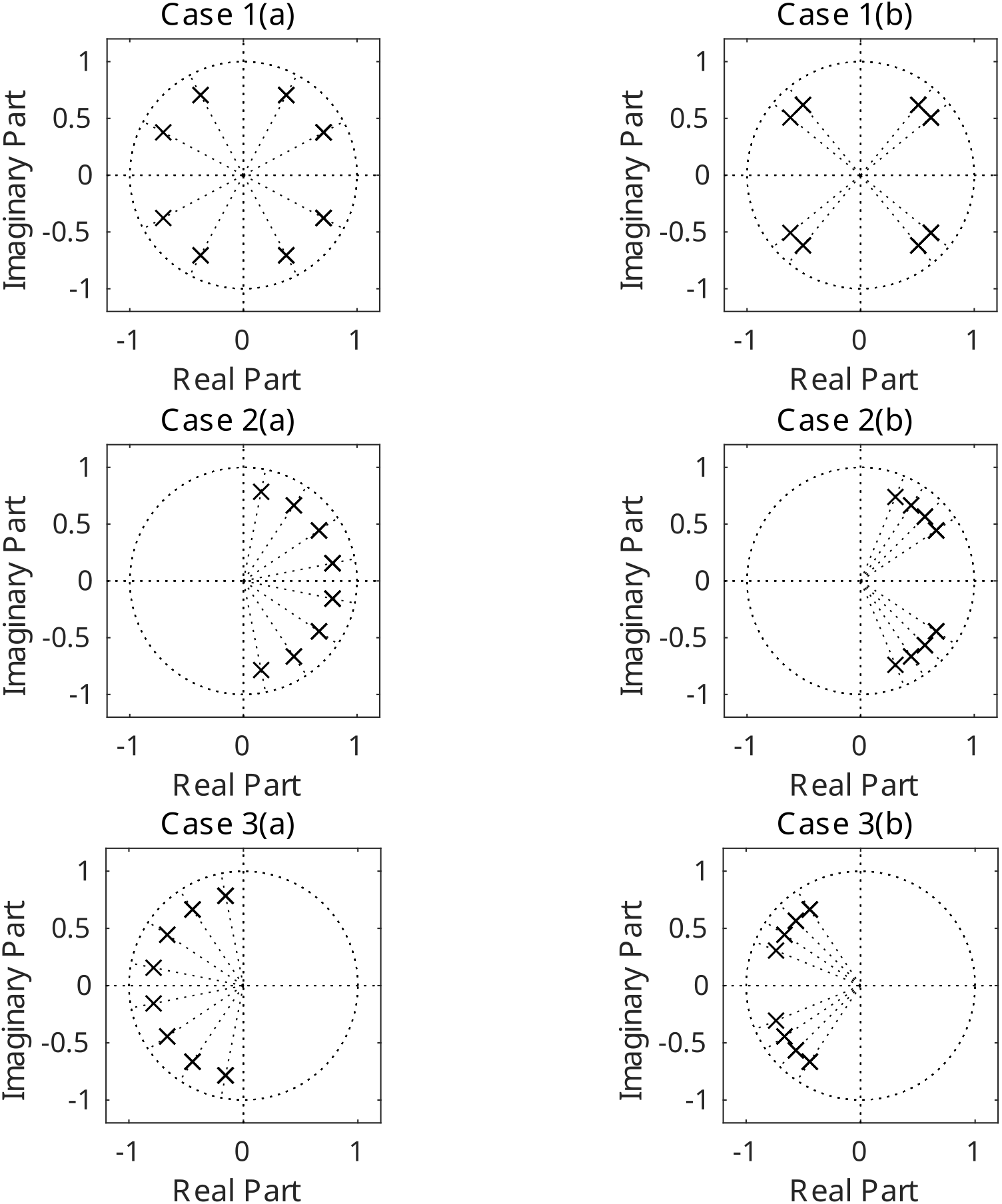}
    \caption{
        The six cases considered for empirical analysis of the error in phase estimation in multi-pole systems.
        As the radii of the poles are varied for each scenario in a particular case, they move along the dotted line.
        }
    \label{fig:zplane_plots}
\end{figure}

\begin{itemize}
    \item The input sequence is generated by the following equation: 
    \begin{equation}
    \label{eq_sinusoid}
        x(n) = \sum_{i=1}^{n_p}a_i^n cos(\omega_in + \phi_i),
    \end{equation}
    for different values of amplitude $a$, angular frequency $\omega$ and phase $\phi$. 
    Here $n_p$ is the number of poles in the signal.
    \item The experiments are broadly divided into six cases based on the position and proximity of the poles.
    They are illustrated in Figure.\ref{fig:zplane_plots} and detailed below:
    \begin{itemize}
        \item Case 1(a) and 1(b) includes two poles in the frequency range (0, $\pi/2$) radians, and two poles in the range ($\pi/2$, $\pi$) radians, that are far apart, and closer to each other, respectively. 
        \item Case 2(a) and 2(b) include four poles in the frequency range (0, $\pi/2$) radians, that are far apart, and closer to each other, respectively. 
        \item Cases 3(a) and 3(b) include four poles in the frequency range ($\pi/2$, $\pi$) radians, that are far apart, and closer to each other, respectively.
    \end{itemize}

    \item For each case, the radius/amplitude $a$ of each pole is varied in increments of 0.05, for $0.7 < a < 1$, to produce numerous scenarios --- more specifically $n_r^{n_p}$ scenarios, where $n_r$ is the number of radii values considered.
    These scenarios consist of poles that are positioned at various locations along the dotted line as illustrated in Figure.\ref{fig:zplane_plots}. 

    \item For each of these scenarios, the chirp spectrum is computed at radius values $0.6 < r_{c} < 1$, in increments of 0.005.
    The chirp spectrum is computed as, 
        \begin{equation}
        \label{eq:FT1}
            X(\omega) = \sum_{n=0}^{N-1} (r_{c}^{-n}x(n))e^{-j\omega n},
        \end{equation} 
    where $r_{c}$ is the radius at which the chirp spectrum is computed.
    The total phase error, is computed each time the chirp spectrum is computed at a particular radius. 
    
    \item The radius at which minimum phase error is obtained for a particular scenario is given by,
        \begin{equation}
            r_{cmin} = arg\ \operatorname*{min}_{r_{c}} 
            \sum_{i=1}^{n_p}(\phi_{p}(i) - \phi_{c}(i,r_{c})),
        \end{equation}
    where $\phi_{p}$ is the actual phase of the pole, and $\phi_{c}$ is the phase computed from the chirp spectrum.
\end{itemize}

\subsubsection{Observations}
\begin{itemize}
    \item Minimum phase error is obtained for all scenarios when $r_{c} = a_{max} + \Delta$, where $a_{max}$ is the radius of the pole with the maximum radius, and $0 < \Delta < 0.1$.

    \item The percentage of the number of scenarios where $\Delta = 0$ is 81\%, and where $\Delta = 0.005$ is 17\%. 
    Hence for almost all scenarios $\Delta \approx 0$.

    \item Considering eq. \ref{eq:FT1}, this is equivalent to computing the chirp spectrum at the radius of the pole with the maximum radius.
    This fact is consistent with the theory explained in Section \ref{sec:one_pole_system}.

    \item It must also be stated explicitly that in no scenario was the minimum phase error found when $r_{c} = a_{max} - \Delta$, for $\Delta > 0$.
    
\end{itemize}

\subsubsection{Conclusions}
\begin{figure}
    \centering
    \includegraphics[width=0.30\textwidth]{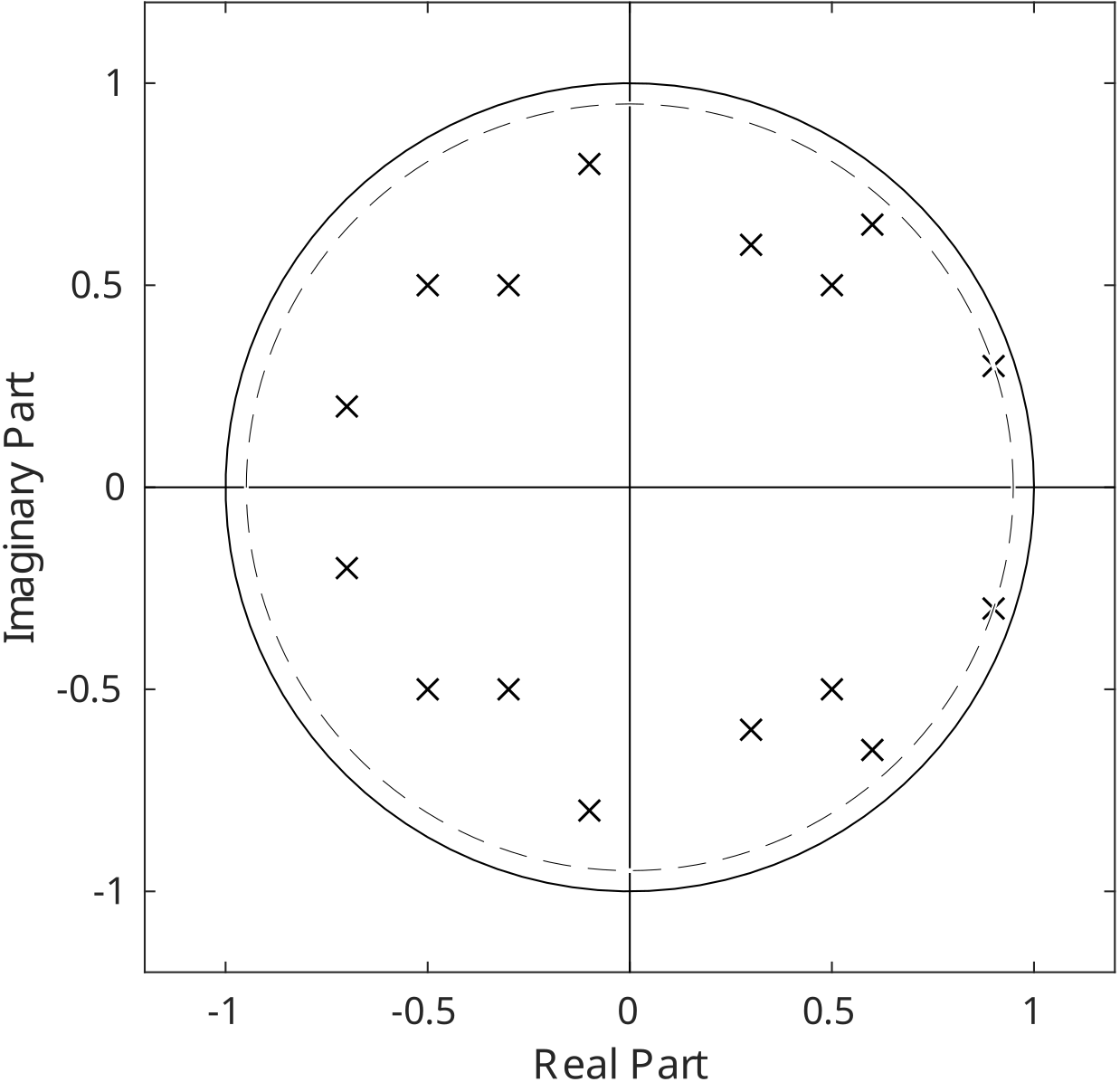}
    \caption{
    Eight complex conjugate poles of a synthesized signal. 
    The solid line marks the unit circle, while the dotted line marks the analysis circle at radius $r_{c}=a_{max}$.}
    \label{fig:zplane-analysis}
\end{figure}

From these observations it can be concluded that the radius $r_{c}$ should be set to $a_{max} + \Delta$ for minimum phase error.
Figure.\ref{fig:zplane-analysis} illustrates this. 
Figure shows a signal with eight poles where the radius of the analysis circle is set to $r_{c}=a_{max}$, or the radius of the pole with the maximum radius.

\subsection{Analysis of Real Speech}
\label{sec:analysis_real_speech}
Real-world \textit{stable} systems can consist of significant poles that are close to, and inside the unit circle ($r<1$), but never on or outside it.
This experiment empirically determines the optimum radius for the computation of the chirp spectrum for real speech signals.

\subsubsection{Experimental Setup}
\begin{itemize}
    \item Totally 400 utterances from the Google speech commands dataset are considered. 
    \begin{itemize}
        \item Each utterance contains one of the 30 speech commands present in the dataset. 
        \item Almost all of these utterances are 1s long and, the sampling rate is 16kHz.
        \item The number of examples, and the duration, of each speech command is near-evenly distributed. 
    \end{itemize}
    \item Linear prediction model is estimated with the order of the filter set to 20.
    \item The location of the poles in each frame, and hence the radius of the pole with the maximum radius is computed.
    \item A histogram of the radius of the pole with the maximum radius across all frames, across all utterances, is plotted in Figure \ref{fig:radius_histogram}.
\end{itemize}

\subsubsection{Observations}
\begin{figure}[ht]
    \centering
    \includegraphics[width=0.48\textwidth]{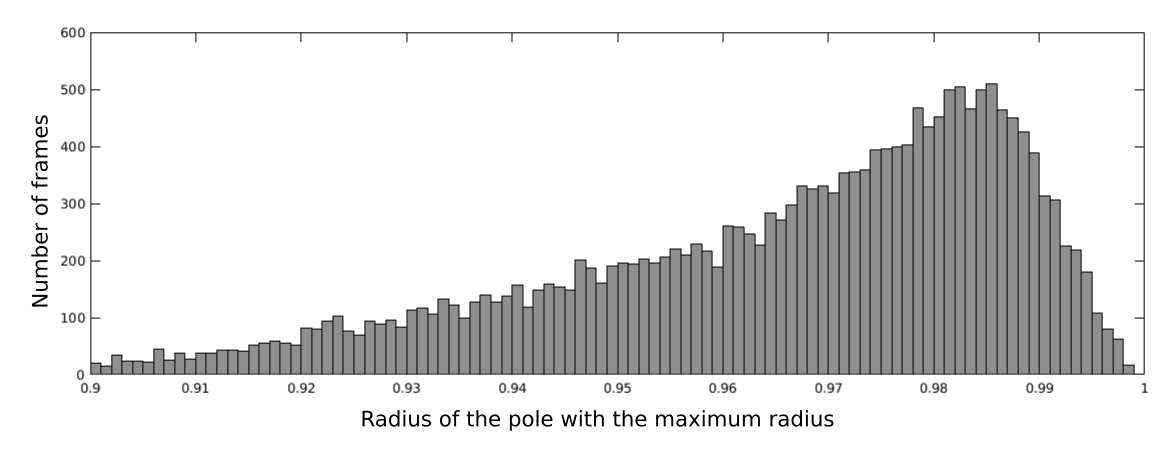}
    \caption{Histogram of the radius of the pole with the maximum radius, computed across 400 (1s long) utterances of the Google speech commands dataset.
    }
    \label{fig:radius_histogram}
\end{figure}

\begin{itemize}
    \item It can be seen from Figure \ref{fig:radius_histogram} that there are no poles at radius $r = 1$, and that the maximum value of $r$ is 0.999.
    \item The histogram rolls off sharply after 0.985, and reaches very minimal values at 0.999. 
\end{itemize}

\subsubsection{Conclusions}
Better estimation of phase, and hence better estimation of the magnitude, can be achieved by, 
\begin{itemize}
    \item assuming $0.997 \leq a_{max} \leq 0.999$ --- a value close to the unit circle but not on it and, 
    \item setting $r_{c} = a_{max}$.
\end{itemize}
Here, coefficient quantization error is assumed to be zero. 
These conclusions are reflected in the computation of the chirp MFCC feature in the experiments that follow.

\subsection{Experimental Comparison of MFCC and Chirp MFCC using POG}
\label{sec:experimental_analysis}

In this section, we use product of Gaussians (POG) \cite{nagarajan2007bias} to compare the class separation offered by MFCC and Chirp MFCC features.
To carry out this experiment, we use five phones --- /aa/, /eh/, /ih/, /ow/, and /uh/ from the TIMIT corpus.

\subsubsection{Chirp MFCC Feature}
Chirp MFCC is derived by computing the chirp spectrum (instead of the discrete Fourier transform in MFCC computation) as in eq \ref{eq_ztransform_2}.
As discussed in the previous sections, assuming a stable system, in which $0 < a_{max} < 1$, the exponential component $a^{-n}$ can be thought of as a multiplier for either the basis function $e^{-j\omega_0n}$ or the input signal $x(n)$.
The algorithm to compute the Chirp MFCC features is as follows.
For each overlapping frame of the signal,
\begin{itemize}
    \item Weigh the frame with the exponentially growing signal $r^{-n}$, of the same size as the frame, where $0 < r < 1$.
    \item Compute the MFCC using this weighted signal.
\end{itemize}
For a specific value of $r \neq 1$, the chirp spectrum is a more accurate representation of the signal.
In the current work, the value of radius $r$ for computing the chirp MFCC is set to $0.990 \leq r \leq 0.999$.

\subsubsection{Experimental Setup}
The experiment can be summarized as follows:
\begin{itemize}
    \item 39 dimensional MFCC (13 static + 13 derivative + 13 acceleration) corresponding to each of the phones are extracted with frame size set to 20ms and frame shift set to 10ms.
    \item For each phone, 3-state hidden Markov models (HMM), with the number of mixture components set based on the population, are built using these features.
    \item 200 examples of each vowel from the test set are tested against all five models, and frame-wise likelihoods are obtained. 
    Here, we assume that the likelihood distribution is ``normal''.
    \item The product of Gaussians, following which the percentage overlap, are computed using the Gaussians obtained from the likelihoods of, 
    \begin{itemize}
        \item a particular phone model (M1) tested with examples of that same phone (P1), and
        \item a particular phone model (M1) tested with examples of a different phone (P2)
    \end{itemize}
        The product of the Gaussians is computed as,
        \begin{equation}
        \begin{split}
            \mu = \frac{\mu_1\sigma_2^2 + \mu_2\sigma_1^2}{\sigma_2^2 + \sigma_1^2} \\
            \sigma^2 = \frac{\sigma_1^2 \sigma_2^2} {\sigma_1^2 + \sigma_2^2}
        \end{split}
        \end{equation}
        where $\mu_1$ and $\mu_2$ are the means of the two Gaussians corresponding to the two sets of likelihoods, $\sigma_1$ and $\sigma_2$ the standard deviation, and $\mu$ and $\sigma$ the mean and standard deviation of the Gaussian product respectively.
        The percentage overlap ($O^{p}$) is given by, 
        \begin{equation}
            O^{p} = e ^{-\left[ {{\frac{\mu - \mu_1}{2\sigma_1^2} + \frac{\mu - \mu_2}{2\sigma_2^2}}} \right]} \times 100.
        \end{equation}
        The percentage overlap computed from the POG (likelihood Gaussians) indicates the similarity (or the class separation), between the two classes. 
        Greater overlap implies more similarity and lesser class separation, hence lower overlap is expected.
    \item All the above steps are repeated for chirp MFCC features as well, and the difference in percentage overlap, when using MFCC and chirp MFCC as a feature, is computed.
    The difference ($D$) between percentage overlap offered by vanilla MFCC $O_{v}^{p}$, and chirp MFCC $O_{c}^{p}$, is measured as, 
    \begin{equation}
        D = O_{v}^{p} - O_{c}^{p}.
    \end{equation}
    Here, positive values of $D$ indicate better class separation offered by chirp MFCC, and vice versa.
\end{itemize}

\begin{figure}[ht]
    \centering
    \includegraphics[width=0.50\textwidth]{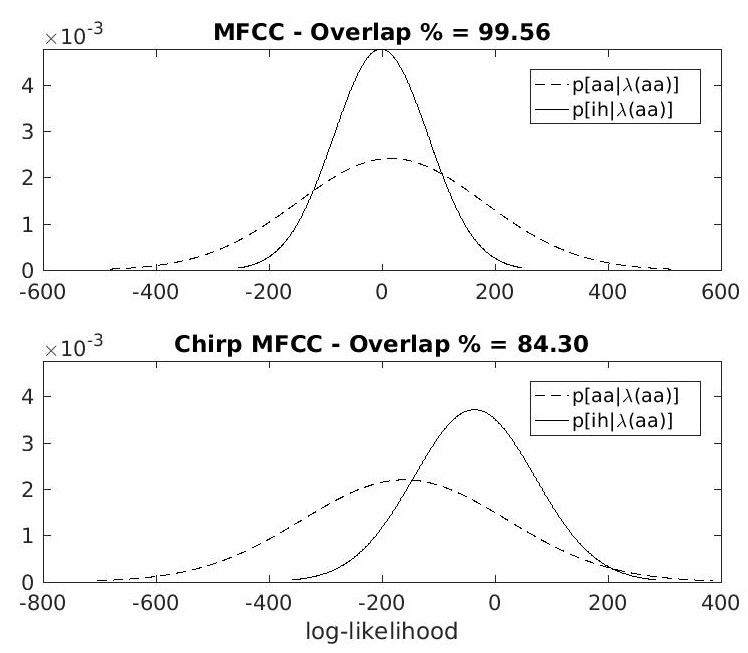}
    \caption{Product of Gaussians showing the difference in percentage overlap offered by MFCC and chirp MFCC.
    The Gaussians correspond to the likelihoods of phone model M1 (/aa/) tested with examples of the same phone P1 (/aa/), and M1 tested with examples of a different phone P2 (/ih/).
    }
    \label{fig:pog}
\end{figure}

Figure. \ref{fig:pog} shows the POG for a single case where phone P1 is /aa/, phone P2 is /ih/ and model M1 is /aa/.
From the figure, it can be seen that for MFCC the overlap is 99.56\%, but for chirp MFCC the overlap is 84.30\%. 
That is, a 15.26\% difference in percentage overlap.
Table. \ref{table:overlap_differences} presents a summary of the differences in percentage overlap for all cases. 

\subsubsection{Observations}
The following observations can be made from Table \ref{table:overlap_differences}:  
\begin{itemize}
    \item There are more positive values than negative values, and the positive values are much higher than the negative values. 
    \item Most of the negative values pertain to one phone (/uh/).
    \item The sum of positive differences is 99.70 (average: 7.67), while the sum of negative differences is 6.01 (average: 0.86).
    \item The last column shows the average percentage overlap difference corresponding to one particular phone model. For all, except /uh/, the average is positive.
    \item The overall mean difference in percentage overlap is 4.68\%.
\end{itemize}

\subsubsection{Conclusion}
From these statistics we can conclude that, by offering better class separation, the contribution of chirp MFCC is predominantly positive with respect to vanilla MFCC.

\begin{table}[]
\centering
\caption{The difference in percentage overlap of likelihood Gaussians, when vanilla MFCC and chirp MFCC are used as a feature. Positive differences show better class separation when using chirp MFCC.}
\label{table:overlap_differences}
\begin{tabular}{c|c|c|c|c|c|c}
\hline \hline
\textbf{M1↓\textbackslash{}P2→} & \textbf{a} & \textbf{e} & \textbf{i} & \textbf{o} & \textbf{u} & \textbf{Average} \\ \hline \hline
\textbf{a}                    &            & 11.31      & 15.26      & -2.62      & 16.97      & 10.23            \\ \hline
\textbf{e}                    & 5.81       &            & 0.01       & 10.11      & -0.51      & 3.86             \\ \hline
\textbf{i}                    & 7.67       & 0.04       &            & 2.31       & -0.81      & 2.30             \\ \hline
\textbf{o}                    & 0.05       & 12.97      & 7.78       &            & 9.41       & 7.55             \\ \hline
\textbf{u}                    & -0.75      & -0.02      & -0.2       &            & -1.1       & -0.52            \\ \hline \hline
\end{tabular}

\end{table}

\section{Evaluation}
\label{sec:evaluation}

\begin{table*}[ht]
\centering
\caption{
    Precision, Recall and F1 measure computed for the Speech, Music and Speech+Music Classification task, 
    using GMM and DNN Classifiers, trained on MFCC and Chirp MFCC features, extracted from 15s and 7.5s test utterances.
    }

\begin{tabular}{lc|ccc|ccc}
\hline \hline
\multirow{2}{*}{\textbf{Classifier}}      & \textbf{Duration →} & \multicolumn{3}{c|}{\textbf{15s Utterances}}                                                 & \multicolumn{3}{c}{\textbf{7.5s Utterances}}                                                 \\ \cline{2-8} 
                                          & \textbf{Feature ↓}  & \multicolumn{1}{c|}{\textbf{Precision}} & \multicolumn{1}{c|}{\textbf{Recall}} & \textbf{F1} & \multicolumn{1}{c|}{\textbf{Precision}} & \multicolumn{1}{c|}{\textbf{Recall}} & \textbf{F1} \\ \hline \hline
\multicolumn{1}{l|}{\multirow{2}{*}{GMM}} & MFCC                & \multicolumn{1}{c|}{0.8744}             & \multicolumn{1}{c|}{0.8556}          & 0.8580      & \multicolumn{1}{c|}{0.8641}             & \multicolumn{1}{c|}{0.8461}          & 0.8490      \\ \cline{2-8} 
\multicolumn{1}{l|}{}                     & Chirp MFCC          & \multicolumn{1}{c|}{0.8874}             & \multicolumn{1}{c|}{0.8722}          & 0.8744      & \multicolumn{1}{c|}{0.8768}             & \multicolumn{1}{c|}{0.8629}          & 0.8648      \\ \hline
\multicolumn{1}{l|}{\multirow{2}{*}{DNN}} & MFCC                & \multicolumn{1}{c|}{0.8709}             & \multicolumn{1}{c|}{0.8611}          & 0.8617      & \multicolumn{1}{c|}{0.8568}             & \multicolumn{1}{c|}{0.8436}          & 0.8441      \\ \cline{2-8} 
\multicolumn{1}{l|}{}                     & Chirp MFCC          & \multicolumn{1}{c|}{0.9099}             & \multicolumn{1}{c|}{0.9056}          & 0.9065      & \multicolumn{1}{c|}{0.8889}             & \multicolumn{1}{c|}{0.8825}          & 0.8836      \\ \hline \hline
\end{tabular}
\label{tab_main}
\end{table*}

To prove the validity of the proposed solution for application to real-world tasks, the following tasks are considered: 
speech-music classification, speaker identification, and speech commands recognition.
For each task, using classifiers relevant to the particular task, two sets of scores are obtained ---
one with vanilla MFCC as the feature and the other with chirp MFCC as the feature.
The difference in performance is then observed, to show whether chirp MFCC offers improvement. 
This is the generic approach followed in each of these tasks, the details are further elaborated in their corresponding sections.

Although the classifiers used for each task are not state of the art classifiers, they are chosen so to make the significance of the feature more evident.
It should also be noticed that the aim of these experimental evaluations is not to demonstrate performance comparable to state-of-the-art systems, but to show the difference between the performance offered by MFCC and chirp MFCC.

\subsection{Speech-Music Classification}
\label{sec:speech-music}
Speech-music classification is the task of classifying an
incoming audio signal into either speech or music.
In the current work, we consider three classes namely, speech, music, and speech+music.
Speech-music classification is generally used as a front-end application which sorts the incoming audio so that the other processing blocks are utilized efficiently, but it has other numerous applications as well
\cite{williams1999speech} \cite{hirvonen2014speech}.

\subsubsection{Approaches in Literature}
The features used for speech-music classification can be either 
handcrafted features \cite{saunders1996real, scheirer1997construction, panagiotakis2005speech, sell2014music, khonglah2016speech} or 
spectral features \cite{mcaulay1986speech, shirazi2010improvement, thoshkahna2006speech}. 
In some cases, as in real-time processing \cite{saunders1996real}, efficiency is key, while in other cases, complexity can be afforded for the sake of accuracy.
Hence selection of one feature over the other is highly task specific \cite{sell2014music}.
An example of a system that employs machine learning techniques for feature extraction can be found in \cite{williams1999speech}.

GMM and DNN are commonly used classifiers in literature.
For example, the use of a multi-variate Gaussian can be found in \cite{saunders1996real} \cite{seck1999two}. 
In \cite{scheirer1997construction}, the use of GMM, kNN, and k-D Trees can be found.
The use of GMM and SVM can be found in \cite{khonglah2016speech} and \cite{hirvonen2014speech}.
The use of MLP Neural Networks can be found in \cite{bugatti2002audio}.

\subsubsection{Dataset}
The dataset used for this task is the music-speech corpus, 
otherwise known as the Scheirer and Slaney (S\&S) dataset \cite{scheirer1997construction}. 
It consists of a set of 15s audio examples recorded from the radio. 
The following are the types of examples present in the dataset: speech (containing speech from both male and female speakers), 
music (music from various genres --- 
with and without vocals),
speech+music (contains speech mixed with music in the background).
`Music with vocals' and `speech + music' are quite similar, but the difference is that vocals/singing is speech with pitch variations.
The training set consists of 120 examples (40 examples from each class) and the testing set consists of 60 examples (20 examples from each class). 
The testing is performed with audio examples with 15s duration as in the corpus, and also with 7.5s duration where the audio examples are split into two.
Three fold cross-validation is carried out on the dataset and the scores are reported.

\subsubsection{Experimental Setup}
In the current work we employ GMM \cite{khonglah2016speech} and DNN \cite{bugatti2002audio} for classification. 
A four mixture component GMM is trained for each class.
The DNN comprises two hidden layers, with 78 nodes each.

\subsubsection{Results and Discussion}
The classification results are summarized in Table \ref{tab_main}.
It can be seen that there is consistent improvement in the F1 scores when chirp MFCC is used instead of vanilla MFCC.
Although the performance of the two classifiers varies (GMM vs DNN), 
the focus of the work is on the improvement that each classifier offers when vanilla MFCC features are replaced by chirp MFCC features.
The improvement offered by chirp MFCC is also consistent when the duration of the test utterances is halved (doubling the number of test utterances).

\subsection{Speaker Identification}
Speaker identification is the process of determining the
speaker's identity given an utterance and a set of speakers.
In the current work, text-independent speaker identification is carried out, where, the features are expected to effectively capture the acoustic characteristics.

\subsubsection{Approaches in Literature}
In literature, GMM \cite{reynolds2000speaker} and 
Factor Analysis \cite{kenny2007joint} \cite{dehak2009support} are commonly used modelling techniques for speaker recognition tasks. 
Joint Factor analysis proposed in \cite{kenny2007joint}, addresses the channel and session variance of a speaker. 
Front-end factor analysis was proposed in \cite{dehak2010front} since joint factor analysis required extensive amount of data to handle channel variations.
Instead of a high-dimensional speaker and channel space, front-end factor analysis defines a single low-dimensional total variability space (TVS).
Here, the speaker dependent supervector $\overrightarrow{M}$
is defined as 

\begin{equation}
    \overrightarrow{M} = \overrightarrow{m} + T\overrightarrow{w},
\end{equation}
where $\overrightarrow{m}$ is the channel- and speaker-
independent supervector, derived from the universal background model (UBM), $T$ is the TVS in low rank space and, $\overrightarrow{w}$ represents the speaker characteristics and is called the identity vector or iVector.

\begin{table}[ht]
\centering
\caption{Precision, Recall and F1 measure computed 
    for the Speaker ID task, 
    using UBM-iVector-based classifier, trained on MFCC and Chirp MFCC features.
    }


\begin{tabular}{l|c|c|c|c}
\hline \hline
\multicolumn{1}{c|}{\textbf{Feature}} & \textbf{Precision} & \textbf{Recall} & \textbf{F1} & \textbf{EER} \\ \hline \hline
MFCC                                  & 0.8160             & 0.8135          & 0.8148      & 7.20         \\ \hline
Chirp MFCC                            & 0.8400             & 0.8254          & 0.8326      & 6.40         \\ \hline \hline
\end{tabular}

\label{tab:speaker_id}
\end{table}

\subsubsection{Dataset}
Voxceleb1 dataset \cite{nagrani2017voxceleb} is used for this task. 
This dataset consists of over 100,000 utterances, 
from 1251 celebrities, extracted from videos uploaded 
to YouTube.
Totally 25 speakers are considered for evaluation,
of which 12 are male and 13 female.
For each speaker, approximately 72\% (105 minutes) of the data is used for training, 11\% (15 minute) for development, and 17\% (24 minutes) for testing.
The mean and standard deviation of the duration of the utterances in train, development, and test are [7.55, 4.24], [6.94, 3.42] and [8.03, 4.67] respectively. 
Hence they are approximately the same length.

\subsubsection{Experimental Setup}
In the current work, we use front-end factor analysis \cite{dehak2010front} and an UBM-iVector-based classifier.
Gender-independent 64 mixture component UBM is built using the training utterances of all the 25 speakers. 
The dimension of i-vectors is 100. 

\subsubsection{Results and Discussion}
The overall precision, recall, F1 measure, and equal error rate (EER) across all the speakers are tabulated in Table \ref{tab:speaker_id}.
It can be seen that there is a 1.8\% improvement in the F1 score, from 0.8148 to 0.8326, when MFCC is replaced with Chirp MFCC as the feature.
There is also a reduction in EER from 7.20 to 6.40.

\subsection{Speech Commands Recognition}
In this task, the system is required to recognize a set of predefined commands/utterances independent of the speaker.

\subsubsection{Approaches in Literature}
Traditionally it is carried out using hidden Markov models (HMM) and more recently with a neural attention model \cite{de2018neural} \cite{rose1990hidden}.
It has applications in mobile devices and other portable computers wherein speech commands can be expected to activate higher level tasks such as automatic speech recognition. 
As a front-end system it reduces computational cost and power consumption \cite{zhang2017hello}.

\subsubsection{Dataset}
For this task, the speech commands dataset by Google \cite{warden2018speech} is utilized. 
It consists of 30 speech commands uttered by 1881 speakers. 
Totally 30 speech commands from 200 speakers are considered for evaluation. 
The train, test, and validation splits are performed using the testing and validation lists provided with the dataset.
On an average 1703 utterances of each speech command is used for training, and 228 utterances for testing.
The utterances are mostly 1s long, with mean utterance duration being 0.98s.

\begin{table}[t]
\centering
\caption{
Precision, Recall and F1 measure computed for the Speech Commands Recognition task, 
using DNN classifier, trained on MFCC and Chirp MFCC features.
}
\begin{tabular}{l|c|c|c}
\hline \hline
\multicolumn{1}{c|}{\textbf{Feature}} & \textbf{Precision} & \textbf{Recall} & \textbf{F1} \\ \hline \hline
MFCC                                  & 0.8240             & 0.8278          & 0.8259      \\ \hline
Chirp MFCC                            & 0.8640             & 0.8354          & 0.8495      \\ \hline \hline
\end{tabular}
\label{tab:speech_commands}
\end{table}

\subsubsection{Experimental Setup}
In the current work, a DNN with two hidden layers with 390 nodes in each layer, is used for recognition.
Both MFCC and chirp MFCC features are extracted for each utterance and they are flattened and fed to the DNN for training and testing.
The precision, recall, and F1 measure are tabulated in Table \ref{tab:speech_commands}.
It can be seen that chirp MFCC offers better performance with an improvement of 2.36\% in the F1 score.

\section{Conclusion}
In this paper, a novel feature, based on the chirp magnitude spectrum --- the chirp MFCC, is proposed. 
It is shown theoretically, and through experimental analysis, that the estimation of the discrete Fourier transform is better when the radius of estimation is, 
(i) the radius of the pole for a single pole system, and 
(ii) the radius of the pole with the maximum radius for a multi-pole system. 
In such a computation, the poles that correspond to decaying oscillations in the signal (tend to) become sustained oscillations, 
and since the basis functions of the discrete Fourier transform are sustained oscillations, it yields a more accurate estimation of the underlying true spectrum. 
The feature has a vast array of applications, simply put: everywhere MFCC is currently bring used, and where finer details in spectrum estimation can lead to better results. 
The application of the proposed feature is demonstrated using three real world tasks where chirp MFCC is found to perform better than vanilla MFCC.


\end{document}